\documentclass[journal=aamick,manuscript=article]{achemso}

\usepackage[version=3]{mhchem} 
\usepackage[T1]{fontenc}       
\usepackage{color}
\usepackage{graphicx}
\usepackage{epstopdf}
\usepackage{mathptmx}
\usepackage{amsmath}

\author{Bin Zhang}
\affiliation{MOE Key Laboratory of Organic OptoElectronics and Molecular Engineering, Department of Chemistry, Tsinghua University, Beijing 100084, China.}
\author{Zhigang Shuai}\email{shuaizhigang@cuhk.edu.cn}
\affiliation{School of Science and Engineering, The Chinese University of Hong Kong, Shenzhen, Guangdong 518172, P R China and MOE Key Laboratory of Organic OptoElectronics and Molecular Engineering, Department of Chemistry, Tsinghua University, Beijing 100084, P R China.}
\title{Quantum Dynamical Approach to Predicting the Optical Pumping Threshold for Lasing in Organic Materials}

\newcommand{\be}{\begin{equation}}
\newcommand{\ee}{\end{equation}}
\newcommand{\ba}{\begin{aligned}}
\newcommand{\ea}{\end{aligned}}
\newcommand{\bea}{\begin{eqnarray}}
\newcommand{\eea}{\end{eqnarray}}

\begin{document}

\begin{abstract}
We present a quantum dynamic study on organic lasing phenomena, which is a challenging
issue in organic optoelectronics. Previously, phenomenological method has achieved success in describing experimental observation. However, it cannot directly bridge the laser threshold with molecular electronic
structure parameters and cavity parameters. Quantum dynamics method for describing organic
lasing and obtaining laser threshold is highly expected. In this Letter, we first propose a microscopic
model suitable for describing the lasing
dynamics of organic molecular system and we apply the time-dependent wave-packet diffusion (TDWPD) to
reveal the microscopic quantum dynamical process for the optical pumped lasing behavior. Lasing threshold is obtained from the onset of output as a function of optical input pumping. We predict that
the lasing threshold has an optimal value as function of the cavity volume and depends linearly
on the intracavity photon leakage rate. The
structure-property relationships between molecular electronic
structure parameters (including the energy of molecular
excited state, the transition dipole and the organization energy) and the laser threshold obtained through numerical calculations are in
qualitative agreement the experimental results, which also confirms the reliability of our approach.
This work is beneficial to understanding the
mechanism of organic laser and optimizing the design of
organic laser materials.
\end{abstract}

\begin{figure}
\centering
\includegraphics[height=7cm]{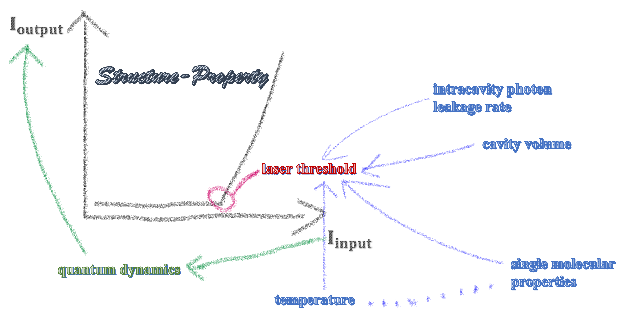}\\

TOC Figure
\end{figure}

The development of organic solid-state lasers (OSSL) have been
greatly promoted over the past few years, due to their
wide-range wavelength and low-cost fabrication.\cite{fichou971178,samuel071272,chenais12390,cui146852,kuehne1612823,gierschner16348} As we know,
organic lasers have been developed for twenty years\cite{jiang205885},
including the optically pumped laser\cite{kuehne1612823,gierschner16348}, the electrically
pumped laser\cite{sandanayaka19061010,ou204485} and the polariton laser\cite{schneider13348,ren207550,jiang212106095}.
At present, the most mature research of OSSLs are the
optically pumped laser. Thanks to the optimization of the
gain medium, the high-Q cavity feedback structure and
the excellent optical excitation system, the performance
of the optically pumped laser has been remarkable improved.\cite{jiang205885} And, the
high-Q optical cavity has became a new way to manipulate the molecular photophysical properties
by light-matter coupling.
Meanwhile, the theoretical design of organic laser molecules
has attracted lots of attention. In the previous work of our group\cite{ou204485,lin22487}, the
computational selection strategies for optically pumped and electrically
pumped organic laser molecules have been proposed.
Compared to the theoretical design of laser molecules,
there are still few researches on the mechanism of
organic lasers, particularly the dynamic process of
intracavity photons.

The formation of laser requires an optical gain to compensate
the photons leakage in a optical cavity.\cite{samuel071272,kuehne1612823,jiang205885}
The laser threshold is a crucial
parameter to describe organic laser performance. The laser threshold
is a particular pump power. When the pump
power is larger than the laser threshold, the output power not only
increases significantly, but also increases linearly with
the input power. Above the laser threshold,
the larger output power for a given input power corresponds to
the larger slope. It is easy to understand
that the lifetime of the exciton and photon in the cavity
is very important for the laser threshold.
And, increasing the lifetime of exciton and photon can reduces
the laser threshold. Excellent
laser performance usually corresponds to low laser threshold. Based on phenomenological theory, Adachi's group
has been studied the influence of different excitonic losses and photonic leakages on the organic laser threshold of organic
lasers under optical and electrical excitations.\cite{yazdani22074003,abe221323} However, the
phenomenological method cannot directly connect between laser
threshold and cavity parameters, molecular electronic
structure parameters. In this paper, we directly relate the input variable and the output variable
of organic lasers to obtain the lasing threshold base on the quantum dynamical method. And then, we investigate the
structure-property relationships between laser threshold and cavity parameters (including intracavity photon
leakage rate, cavity volume), molecular electronic structure parameters (including the energy of molecular
excited state, the transition dipole and the organization energy). Finally, we obtain the physical picture of
organic lasers and the related structure-property relationships. The outline of this paper is as follows: First,
we extend the TDWPD\cite{zhong11134110,zhong13014111,han15578,zhang211654} method including the light-matter
interaction for describing the organic laser in dissipative cavity. Then,
combining the extended TDWPD coupled with properly electronic structure
calculations of 4,4$'$-bis[(N-carbazole)styryl]biphenyl (BSBCz)\cite{sandanayaka16834} to investigate the structure-property relationship between the
intracavity photon leakage rate, cavity volume, single molecular
electronic structure properties and the laser threshold. Finally, we
investigate the influence of the temperature and the external field
duration on the laser threshold. The proposed formalism and the
structure-property relationship is beneficial to understanding the
mechanism of organic laser and optimizing the design of
organic laser materials.
\begin{figure}[ht]
 \centering
  \includegraphics [height=5cm,angle=0]{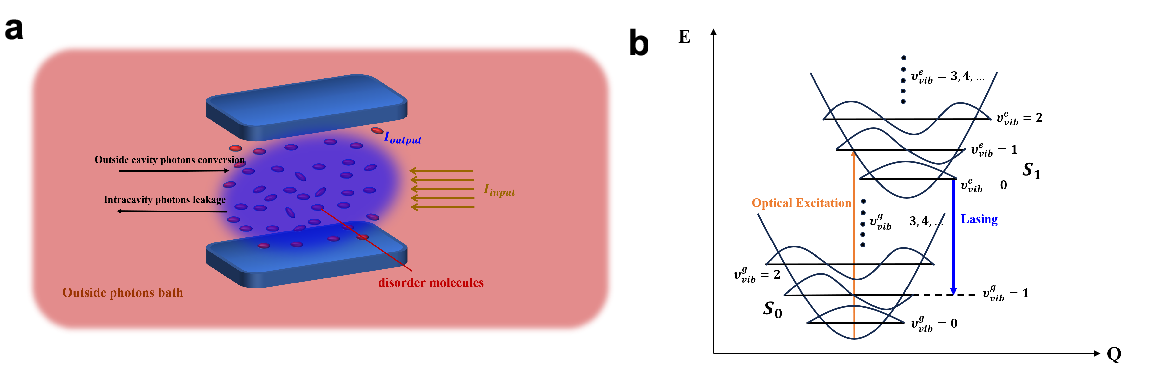}
  \caption{Panel a: Schematic graph of organic lasers in dissipative cavity; Panel b: The molecular four-level energy system, where $E$ is energy, $Q$ is the vibrational coordinate.}
  \label{fig1}
\end{figure}

In this paper, we study the system which is $N$ identical molecules
inside a dissipative optical cavity. The interaction strength between the ith
molecule and the intracavity photon $\hbar g_i$ can be written as\cite{ou2117786,ebbesen162403}
\bea
\hbar g_i=|\vec{\mu}_{eg}|\sqrt{\frac{\hbar\omega_P}{2\epsilon_0\epsilon_{\infty}V}}\cos\theta_i \label{eq1}
\eea
where $\vec{\mu}_{eg}$ is the transition dipole moment (TDM) of the $S_1$ state; $\omega_P$
is the frequency of the intracavity photon; $\epsilon_0$ is the vacuum
permittivity; $\epsilon_{\infty}$ is the optical dielectric constant
of the matrix inside the cavity; $V$ is the cavity mode volume; and
$\theta_i$ is the angle between the TDM of the ith $S_1$ and the
intracavity photon. Within the random orientation model,
disorder molecules are independent of each other. So, the
lasing process of each molecule can be independently studied. And,
we assume that each effective molecule
only couples one intracavity photon, and they compose a subsystem that is also independent each other. And, the effective molecular
number is $N_{eff}=N/\sqrt{3}$. So, we
can use the quantum dynamics method to calculate
the dynamical properties inside the subsystem including one-molecule and one-photon
and then obtain the properties of the total intracavity system. The
effective coupling $\hbar\bar{g}$ can be expressed as
\bea
\hbar\bar{g}=\frac{1}{\sqrt{3}}|\vec{\mu}_{eg}|\sqrt{\frac{\hbar\omega_p}{2\epsilon_0\epsilon_{\infty}V}} \label{eq2}
\eea
In the following text, we use the cavity length $L_{cavity}$ to replace the cavity
volume $V=L_{cavity}^3$. And, we use $V_{eP}$ to replace the
effective exciton-photon coupling $\hbar\bar{g}$.
To describe the organic lasing dynamics in dissipative cavity,
the total Hamiltonian can be expressed as
\be
\ba
\hat{H}(t)&=\hat{H}_e(t)+\hat{H}_{photon}+\hat{H}_{phonon}+\hat{H}_{loss}+\hat{H}_{e-photon}+\hat{H}_{e-phonon}+\hat{H}_{photon-loss} \\
&=\hat{H}_E(t)+\hat{H}_{E-bath}+\hat{H}_{bath} \label{eq3}
\ea
\ee
Here,~$\hat{H}_e(t)$, $\hat{H}_{phonon}$, $\hat{H}_{e-phonon}$, $\hat{H}_{photon}$, $\hat{H}_{e-photon}$,
$\hat{H}_{loss}$ and $\hat{H}_{photon-loss}$ denote the Hamiltonian for exciton,
the vibrations (or phonon), the exciton-phonon couplings, the intracavity photon,
the light-matter interaction, the outside cavity bath and the intracavity
photon-outside cavity bath coupling, respectively. And,~$\hat{H}_E(t)=\hat{H}_e(t)+\hat{H}_{photon}+\hat{H}_{e-photon}$,
~$\hat{H}_{E-bath}=\hat{H}_{e-phonon}+\hat{H}_{photon-loss}$ and $\hat{H}_{bath}=\hat{H}_{phonon}+\hat{H}_{loss}$. The
Hamiltonian inside the intracavity subsystem including
one-molecule and one-photon can be written as
\be
\left\{
\ba
&\hat{H}_e(t)=\epsilon_g|g\rangle\langle g|+\epsilon_e|e\rangle\langle e|-\hat{\vec{\mu}}_{eg}\cdot\vec{E}_{pump}(t)\\
&\hat{H}_{photon}=\hbar\omega_{P}(\hat{c}_{P}^{\dagger}\hat{c}_{P}+\frac{1}{2})\\
&\hat{H}_{phonon}=\sum_{j=1}^{N_{ph}^{e}}\hbar\omega_{j}^{e}(\hat{a}_{ej}^{\dagger}\hat{a}_{ej}+\frac{1}{2})\\
&\hat{H}_{loss}=\sum_{j=1}^{N_{b}^{P}}\hbar\omega_{j}^{P}(\hat{b}_{Pj}^{\dagger}\hat{b}_{Pj}+\frac{1}{2})\\
&\hat{H}_{E-photon}=V_{eP}|e\rangle\langle g|\hat{c}_{P}+h.c.\\
&\hat{H}_{E-phonon}=\sum_{j=1}^{N_{ph}^{e}}C_{j}^{e}(\hat{a}_{ej}^{\dagger}+\hat{a}_{ej})|e\rangle\langle e|\\
&\hat{H}_{photon-loss}=\sum_{j=1}^{N_{b}^{P}}C_{j}^{P}(\hat{b}_{Pj}^{\dagger}+\hat{b}_{Pj})(\hat{c}_{P}^{\dagger}+\hat{c}_{P})\label{eq4}
\ea
\right.
\ee
 And, $\epsilon_g$, $\omega_P$
and $\epsilon_e$ is the energy of ground state $|g\rangle$,
the photonic state $\hat{c}_{P}^{\dagger}|g\rangle$ and the localized
singlet excited state $|e\rangle$, respectively. In this work, the phonon and the outside
cavity bath are identified by a collection of harmonic oscillators.~$C_{ej}$~is
the mode-specific electron-vibrational coupling strength, and it is determined by the spectral density
~$J_e(\omega)=\pi\sum_j C_{ej}^2\delta(\omega-\omega_{j}^{e})$.~$C_{j}^{P}$~represents the
intracavity photon-outside cavity bath coupling strength, and it is determined by
the spectral density~$J_P(\omega)=\pi\sum_j {C_{j}^{P}}^2\delta(\omega-\omega_{j}^{P})$. Here, we
set $\vec{\mu}_{eg}\cdot\vec{E}_{pump}(t)\approx |\mu_{eg}|E_{pump}(t)/\sqrt{3}$ which is also
the result based on the random orientation approximation. The external field $E_{pump}(t)$ is
$E_{pump}(t)=\frac{E_0}{\sqrt{2\pi}\sigma}e^{-\frac{t^2}{2\sigma^2}}\cos(\omega_{pump}t)$. And,~$\sigma$,~$E_0$~
and~$\omega_{pump}$~are the field duration, the field strength and the field frequency, respectively. In
the following text, we use $\mu_{eg}$ to replace $|\mu_{eg}|$. For resonance excitation, we
set $\hbar\omega_{pump}=\epsilon_e$. The formulas and
the parameters setting of their spectral density will be described in
detail in the following text. In the quantum dynamic simulations, we adopt TDWPD~\cite{zhong11134110,zhong13014111,han15578}
because it can be easily applied to large complex systems and extended to incorporate the strong
light-matter coupling. Recently, TDWPD method has been extended to
to include the
light-matter interaction and successfully investigated the effect of the optical microcavity on
the singlet fission dynamics in organic systems.\cite{zhang211654} Previous research results
have shown that the TDWPD method is indeed suitable for studying the dynamic properties
of complex molecular systems incorporating light-matter interaction. The TDWPD method
is one of stochastic Schr$\ddot{o}$dinger equations(SSE)
where the molecular vibrational motions are described by random
fluctuations on each electronic state. And, the dynamical equation can be written as
\bea
i\frac{\partial}{\partial t}|\Psi(t)\rangle=(\hat{H}_E(t)+\hat{F}(t)-i\hat{L}\int_{0}^{t}d\tau\alpha_{T=0}(\tau)e^{i\int_{0}^{\tau}\hat{H}_E(\tau^{'})d\tau^{'}}\hat{L}^{\dagger}e^{-i\int_{0}^{\tau}\hat{H}_E(\tau^{'})d\tau^{'}})|\Psi(t)\rangle \label{eq5}
\eea
where $\hat{H}_E(t)$ is the intracavity system Hamiltonian, $\hat{F}(t)$ is the
the stochastic force operator $\hat{F}(t)=\sum_{n,m}F_{nm}(t)|n\rangle\langle m|$, $\alpha_{T=0}(t)$ is the zero-temperature
correlation function $\alpha_{T=0}(t)=\sum_{j}{C_{j}^{n}}^2e^{-i\omega_{j}^{n}t}$, $\hat{L}$ is
the projection operator. The states $|n(m)\rangle$ include the ground state $|g\rangle$, the excited
state $|e\rangle$ and the intracavity photon state $|P\rangle$. And,
$\hat{L}=|e\rangle\langle e|$, $|g\rangle\langle P|$ and $|P\rangle\langle g|$
are correspond to the excited state $|e\rangle$, the ground state $|g\rangle$
and the intracavity photon state$|P\rangle$, respectively. The completely
relation is
\bea
1=|e\rangle\langle e|+|g\rangle\langle g|+|P\rangle\langle P| \label{eq6}
\eea
In numerical
calculations, the wavefunction $|\Psi(t)\rangle$ is written as
\bea
|\Psi(t)\rangle=A_e(t)|e\rangle+A_g(t)|g\rangle+A_P(t)|P\rangle \label{eq7}
\eea
The differential equations of the time-dependent coefficients $\{A_j(t)\}$ (j=g, e or P) are
\be
\label{eq8}
\left\{
\ba
i\frac{\partial}{\partial t}A_e(t)=&(\epsilon_e+F_e(t))A_e(t)+V_{eP}A_P(t)-(\vec{\mu}_{eg}\cdot\vec{E}_{pump}(t))A_g(t)\\
&-i\sum_{k}\int^t_0d\tau\alpha_e(\tau)\langle e|e^{i\int_{0}^{\tau}\hat{H}_E(\tau^{'})d\tau^{'}}|e\rangle\langle e|e^{-i\int_{0}^{\tau}\hat{H}_E(\tau^{'})d\tau^{'}}|k\rangle A_k(t),\\
i\frac{\partial}{\partial t}A_g(t)=&\epsilon_gA_g(t)+F_g(t)A_P(t)-(\vec{\mu}_{eg}\cdot\vec{E}_{pump}(t))A_e(t)\\
&-i\sum_{k}\int^t_0d\tau\alpha_g(\tau)\langle P|e^{i\int_{0}^{\tau}\hat{H}_E(\tau^{'})d\tau^{'}}|P\rangle\langle g|e^{-i\int_{0}^{\tau}\hat{H}_E(\tau^{'})d\tau^{'}}|k\rangle A_k(t),\\
i\frac{\partial}{\partial t}A_P(t)=&\hbar\omega_{P} A_P(t)+V_{eP}A_e(t)+F_P(t)A_g(t)\\
&-i\sum_{k}\int^t_0d\tau\alpha_P(\tau)\langle g|e^{i\int_{0}^{\tau}\hat{H}_E(\tau^{'})d\tau^{'}}|g\rangle\langle P|e^{-i\int_{0}^{\tau}\hat{H}_E(\tau^{'})d\tau^{'}}|k\rangle A_k(t).
\ea
\right.
\ee
Here, $F_{e(P)}(t)$~is the stochastic force, which can be generated by
\bea
F_{e(P)}(t)=&\sum_k\sqrt{\frac{J_{e(P)}(\omega_k)\triangle\omega}{\pi}}[\sqrt{A(\omega_k)}\cos(\omega_k t+\phi_k)+i\sqrt{B(\omega_k)}\sin(\omega_k t+\phi_k)] \label{eq9}
\eea
with~$A(\omega_k)=\coth(\omega_k/2k_BT)+csch(\omega_k/2k_BT)$ and~$B(\omega_k)=\coth(\omega_k/2k_BT)-csch(\omega_k/2k_BT)$.
$\{\phi_k\}$ is
a series of random variables that are uniformly distributed in $[0,2\pi]$.~$\alpha_{e}(t)=\sum_{j}{C_{j}^{e}}^2e^{-i\omega_{j}^{e}t}$, $\alpha_{P}(t)=\sum_{j}{C_{j}^{P}}^2e^{-i\omega_{j}^{P}t}$~are
the zero-temperature correlation function of the exciton-phonon
couplings and the intracavity photon-outside cavity bath couplings. From the TDWPD equation Eq. \ref{eq8}, we can obtain
the time-dependent coefficients $\{A_j(t)\}$ (j=g, e or P). The population dynamics can be obtained
by the stochastic average of $A_{e(g,~or~P)}(t)$. For instance, the time
evolution of population on the $i(i=e,~g~or~P)$-th state is calculated by $P_i(t) = <A_i^*(t) A_i(t)>$. The
input variable is the intensity of incident laser pulse
is $I_{input}$, and it can be defined as
\bea
I_{input}\propto|E_0|^2 \label{eq10}
\eea
where $1.0$~a.u. field strength$|E_0|$~corresponds to $3.5094\times 10^{13}$~KW/cm$^2$ input
intensity $I_{input}$.\cite{sun07195438,sun20224708} Although $I_{input}$ is
the input variable, we directly change the field strength $E_0$ in
numerical calculations. The output variable is $I_{output}$, which is the steady-state photon density
inside the cavity, and it can be defined as
\bea
I_{output}=\lim_{t\rightarrow\infty}P_P(t)c_MN_A/\sqrt{3} \label{eq11}
\eea
where $P_P(t)$ is the population of photon state inside the intracavity subsystem including one-molecule
and one-photon, $c_M=N/V$ is the doping concentration of the molecule,
$N_A$ is the Avogadro constant. $\lim_{t\rightarrow\infty}P_P(t)$ is the population of the intracavity photon
state when the intracavity system reaches the thermal equilibrium state. It is noteworthy that we should use the effective molecular
number $N_{eff}=N/\sqrt{3}$ to calculate the output variable $I_{output}$ due to the random
orientation approximation.\cite{ou2117786} For the intracavity system including exciton,
phonon, and photon, we use different
intensities of external fields $I_{input}$ to excite molecules and transfer the
population to the intracavity photon state through exciton-intracavity photon coupling. When
the steady-state state is reached, the intracavity photon density is
calculated to obtain the output variable $I_{output}$.\\

\begin{figure}[ht]
 \centering
  \includegraphics [height=8cm,angle=0]{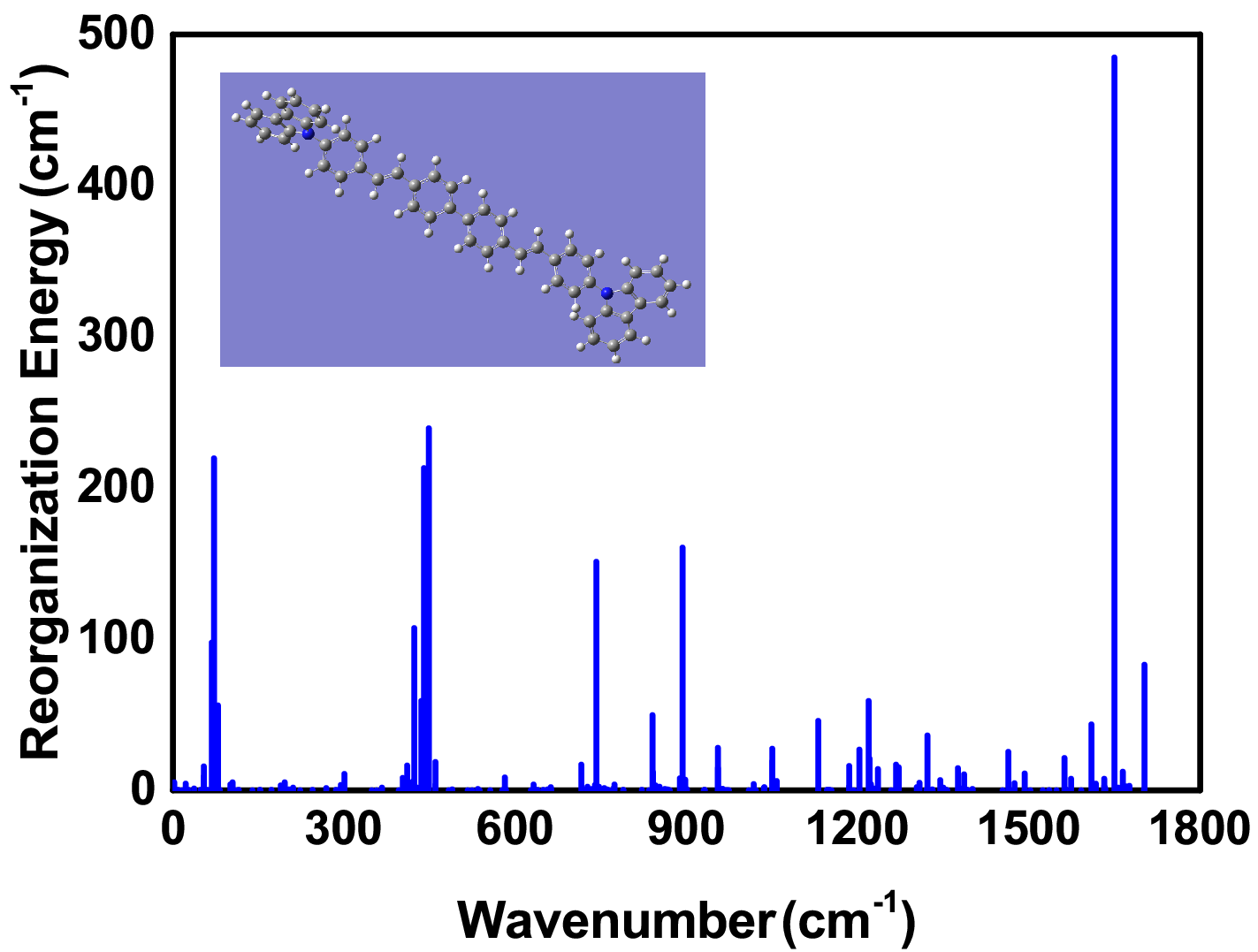}
  \caption{The reorganization energy distribution of BSBCz. The inset shows the chemical structure of BSBCz.}
  \label{fig2}
\end{figure}

\textbf{The Molecular Properties of BSBCz.}
In this paper, we use BSBCz as a test molecule. In the previous work of our
group, BSBCz not only has excellent the photo-pumped laser performance, but
also has excellent the electro-pumped laser performance.\cite{ou204485} We use TD-B3LYP/6-31g*
to calculate the single molecule properties of BSBCz by Gaussian16\cite{g16}. Theoretically
predicted that the energy of $S_1$ state is $2.472$~eV, and the transition dipole is $6.839$~a.u.. Based on
the knowledge of electronic structure, we compute the
rate constants of different physical processes by the TVCF rate
formalism.\cite{lin212944,wang2018721,shuai14123} All rate constant
calculations and are performed via thermal vibration correlation function
(TVCF) method in MOMAP 2021A.\cite{shuai201223,shuai17224,peng079333} Theoretically
predicted that the radiative rate of $S_1$ state $k_r$ is $6.5\times 10^8$~s$^{-1}$ as well
as the internal conversion rate of $S_1$ state $k_{ic}$ is $1.5\times 10^8$~s$^{-1}$, which
are consistent with the numerical results in Ref. \citenum{ou204485}. The Lorentzian
broadening $FWHM~=~200$~cm$^{-1}$ has been used to $k_{ic}$ for converge. Here,
we also display the molecular reorganization energy distribution, shown in
Figure \ref{fig2}. In the following text, we get the intramolecular
vibration spectral density based
on the reorganization energy distribution. Our approach cannot does not destroy the molecular
four-level energy system (see Figure \ref{fig1} b) due to including all of vibration information.\\

\textbf{The Quantum Dynamics Results of Organic Lasers.}
Next, we calculate the laser dynamics, as well as the structure-property relationship
between the intracavity photon-outside cavity bath coupling strength, cavity size, single molecular
electronic structure properties and the laser threshold. Before
the numerical calculations, we summarize the extended TDWPD method and the calculation method of the
main physical quantities involved in this Letter. In this paper, we study the system which is $N$ identical molecules
inside an dissipative optical cavity. Based on the random
orientation approximation, the disorder molecules inside the cavity are independent of each other. The
lasing process of each molecule can be independently studied. And, we assume that each effective molecule
only couples one intracavity photon, and they compose a subsystem that is also independent each other. So, we
can use the quantum dynamics method to calculate
the dynamical properties inside the subsystem including one-molecule and one-photon and then obtain the
properties of the total intracavity system. And, the effective molecular
number is $N_{eff}=N/\sqrt{3}$. In the quantum dynamic simulations, we adopt TDWPD~\cite{zhong11134110,zhong13014111,han15578}
because it can be easily applied to large systems and extended to to incorporate the strong
light-matter coupling. Recently, TDWPD method has been extended to
to include the
light-matter interaction and successfully investigated the effect of the optical microcavity on
the singlet fission dynamics in organic system.\cite{zhang211654} Previous research results
have shown that the TDWPD method is indeed suitable for studying the dynamic properties
of complex molecular systems incorporating light-matter interaction. The TDWPD method
is one of stochastic Schr$\ddot{o}$dinger equations(SSE) where the molecular vibrational motions are described by random
fluctuations on each electronic state. From the TDWPD equation Eq. \ref{eq8}, we can obtain
the time-dependent coefficients $\{A_j(t)\}$ (j=g, e or P). The population dynamics is thus obtained
by the stochastic average of $A_{e(g,~or~P)}(t)$. For example, the time
evolution of population on the $i(i=e,~g~or~P)$-th state is calculated by $P_i(t) = <A_i^*(t) A_i(t)>$. The input variable
is $I_{input}$ ($I_{input}\propto|E_0|^2$, see Eq. \ref{eq10}) is the
input variable is the intensity of incident laser pulse $E_{pump}(t)$, and we change the intensity of incident laser
pulse by changing the value of the field strength $E_0$ in the numerical calculations. The output
variable $I_{output}$ ($I_{output}=\lim_{t\rightarrow\infty}P_P(t)c_MN_A/\sqrt{3}$, see Eq. \ref{eq11}) is the
steady-state photon density inside the cavity. And,
the doping concentration $c_M$ is $0.15\times 10^{-3}$~M.\cite{sandanayaka16834} Next, we can obtain the laser
threshold $I_{thred}$ and the field strength threshold $E_{thred}$ through the inflection point of the
$I_{output}$-$I_{input}$ curve. Following the preceding calculation process, we can obtain the structure-property
relationships between laser threshold $I_{thred}$ and cavity parameters, molecular electronic structure parameters.

For our microscopic
model, the coupling between the intracavity photons and bath is actually the coupling between the intracavity photons and
the continuous photon environment outside the cavity. It leads to the intracavity photons leak into
the photon environment outside the cavity, manifested as the quenching of the intracavity photons.
Similarly, if there are no intracavity photons, and the intracavity photons-bath coupling is relatively large,
the photons outside the cavity will also penetrate into the cavity (at the initial time, the initial
state of the photon environment is thermal equilibrium distribution). So, there will be a small amount
of population of the intracavity photon state without external field $E_{pump}(t)$ excitation. The schematic graph
is shown as Figure \ref{fig1}. The population dynamics of the intracavity system at exciton-photon
coupling $V_{eP}$ also confirms the physical picture, shown in Figure S5, S6 (see the Supporting Information). For
exciton-phonon coupling~$C_{ej}$, is described by the
spectral density with broadened stick-spectra of pseudo-local phonon
modes\cite{berkelbach13114103,zang175105}, $J_e(\omega)=\frac{1}{\pi}\sum_k\frac{\lambda_k\omega\gamma}{(\omega-\omega_k)^2+\gamma^2}$ which
includes the information of all vibrational modes to study the dynamical
properties of the realistic molecular system with a uniform broadening factor of
$\gamma=40.0$~meV to smoothly generate fluctuation energies.\cite{zhu1722587} The reorganization energy ${\lambda_k}$
is calculated via the vibrational modes of the monomer, shown in Figure \ref{fig2}. The coupling
strength~$C_{j}^{P}$~between intracavity photon and we use
the Debye spectral density $J_P(\omega)=\frac{2\lambda_P\omega\omega^c_P}{\omega^2+{\omega^{c}_P}^2}$ to describe bath,
which is consistent with David Reichman's recent work\cite{lindoy232733}. And, we set the characteristic frequency
of the photon environment outside the cavity $\omega^c_P$ is ~$\omega^c_P=1450$~cm$^{-1}$. According to the
spectral density $J_P(\omega)$, we can calculate the corresponding leakage rate of the intracavity
photon $\Gamma_{P}=\frac{1}{\tau_P}=\frac{2J_{P}(\omega_P)}{1-e^{-\beta\omega_P}}$ where $\tau_c$ is the intracavity photon
lifetime and $\beta=\frac{1}{k_BT}$ with the Boltzmann constant $k_B$ and temperature T($T=300~K$), and then obtain
the quality of cavity(Q value) $Q=\frac{1}{2}\frac{\omega_P}{\Gamma_{P}}$.\cite{yazdani22074003,abe221323} The relevant results are
shown in the Table \ref{tab1}. The exciton-photon coupling strength is shown in the Table \ref{tab3}.
In this paper, we set the resonance between the external field energy
and the $S_1$ state of molecule. From the Figure \ref{fig7}, we can see that the duration $\sigma$ is almost
independent of the laser threshold. Without loss of generality, we set the $\sigma$ to be $30.0$~fs later.
\begin{figure}[ht]
 \centering
  \includegraphics [height=6cm,angle=0]{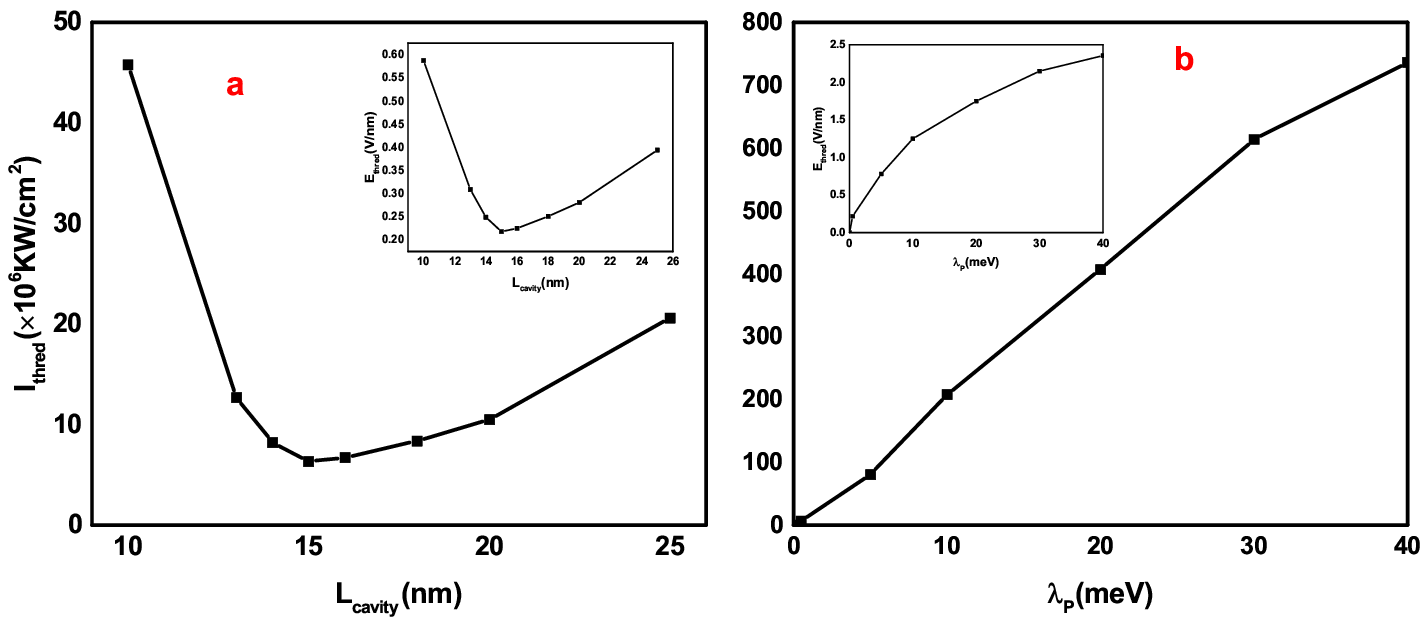}
  \caption{The structure-property relationship between cavity length $L_{cavity}$ (Panel a), the intracavity photon-outside
cavity bath coupling strength $\lambda_P$ (Panel b),  and laser threshold $I_{thred}$. The inset shows the structure-property
relationship about the field strength threshold $E_{thred}$.}
  \label{fig3}
\end{figure}

\begin{table}[ht]
  \renewcommand\arraystretch{1.5}
  \caption{The intracavity photon leakage rate $\Gamma_P$ and the quality of cavity $Q$ to different intracavity photon-outside
cavity bath coupling strength $\lambda_P$.}
  \medskip
  \begin{tabular}{cccccccccc}
  \hline
  \hline
     $\lambda_P (meV)$       &0.0         &0.5      &1.0      &5.0     &10.0    &20.0    &50.0   &100.0  &200.0 \\
     $\Gamma_P (meV)$        &0.0         &0.1447   &0.289    &1.447   &2.894   &5.787   &14.47  &28.94  &57.87 \\
     Q                       &infinity    &8542.64  &4271.32  &854.26  &427.13  &213.57  &85.43  &42.71  &21.36 \\
  \hline
  \end{tabular}
  \label{tab1}
\end{table}

\begin{table}[ht]
  \renewcommand\arraystretch{1.5}
  \caption{The intracavity photon leakage rate $\Gamma_P$ and the quality of cavity $Q$ to different intracavity photon frequency $\omega_P$.}
  \medskip
  \begin{tabular}{ccccccc}
  \hline
  \hline
     $\omega_P (eV)$        &2.0              &2.2           &2.4            &2.6          &2.8         &3.0 \\
     $\Gamma_P (meV)$       &0.1783           &0.1624        &0.1490         &0.1376       &0.1279      &0.1190 \\
     Q                      &5607.37          &6775.48       &8054.84        &9445.44      &10947.30    &12560.40 \\
  \hline
  \end{tabular}
  \label{tab2}
\end{table}

\begin{table}[ht]
  \renewcommand\arraystretch{1.5}
  \caption{The exciton-photon coupling $V_{ep}$ to different cavity length $L_{cavity}$.}
  \medskip
  \begin{tabular}{cccccccc}
  \hline
  \hline
     $L_{cavity} (nm)$       &10.0     &13.0    &15.0   &16.0   &18.0   &20.0   &25.0 \\
     $V_{ep} (meV)$          &31.55    &21.08   &17.18  &15.44  &12.94  &11.16  &7.91 \\
  \hline
  \end{tabular}
  \label{tab3}
\end{table}

\begin{figure}[ht]
 \centering
  \includegraphics [height=10cm,angle=0]{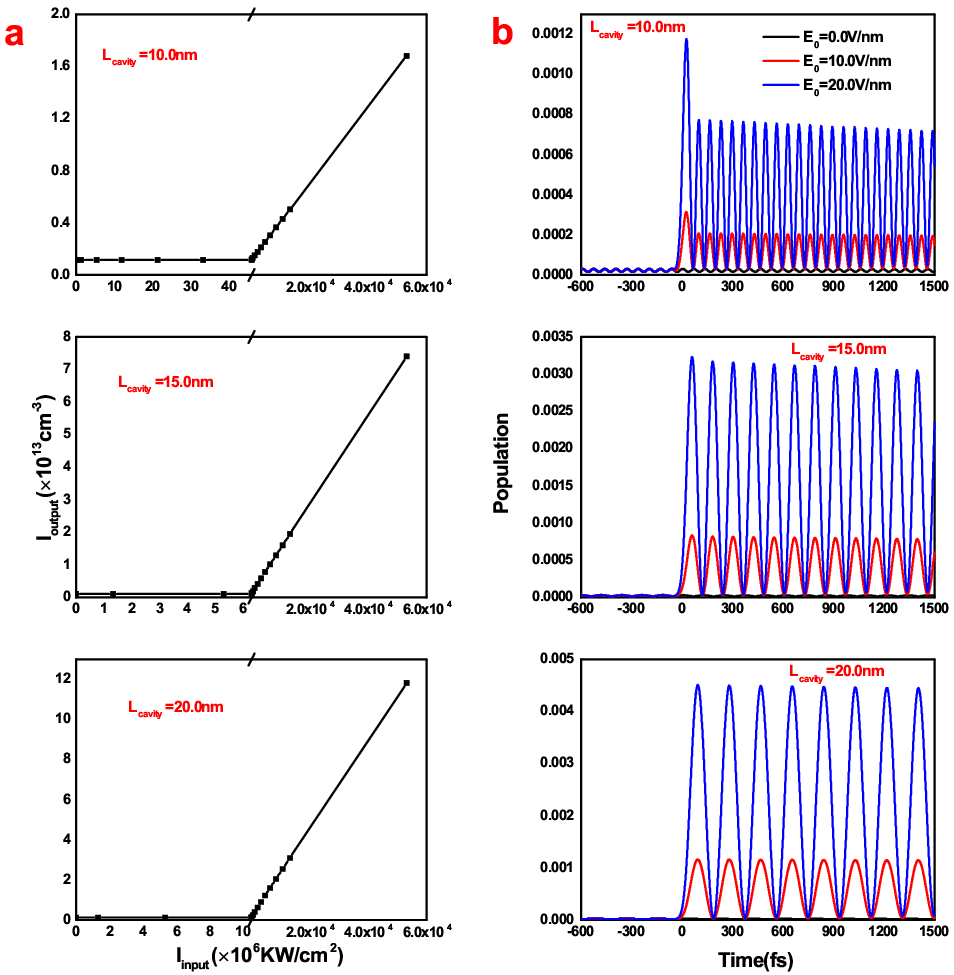}
  \caption{Panel a: The $I_{output}$-$I_{input}$ curve at different cavity length $L_{cavity}$; Panel b: The population dynamics of photon state inside the subsystem including one-molecule and one-photon at different cavity length $L_{cavity}$.}
  \label{fig4}
\end{figure}

\begin{figure}[ht]
 \centering
  \includegraphics [height=10cm,angle=0]{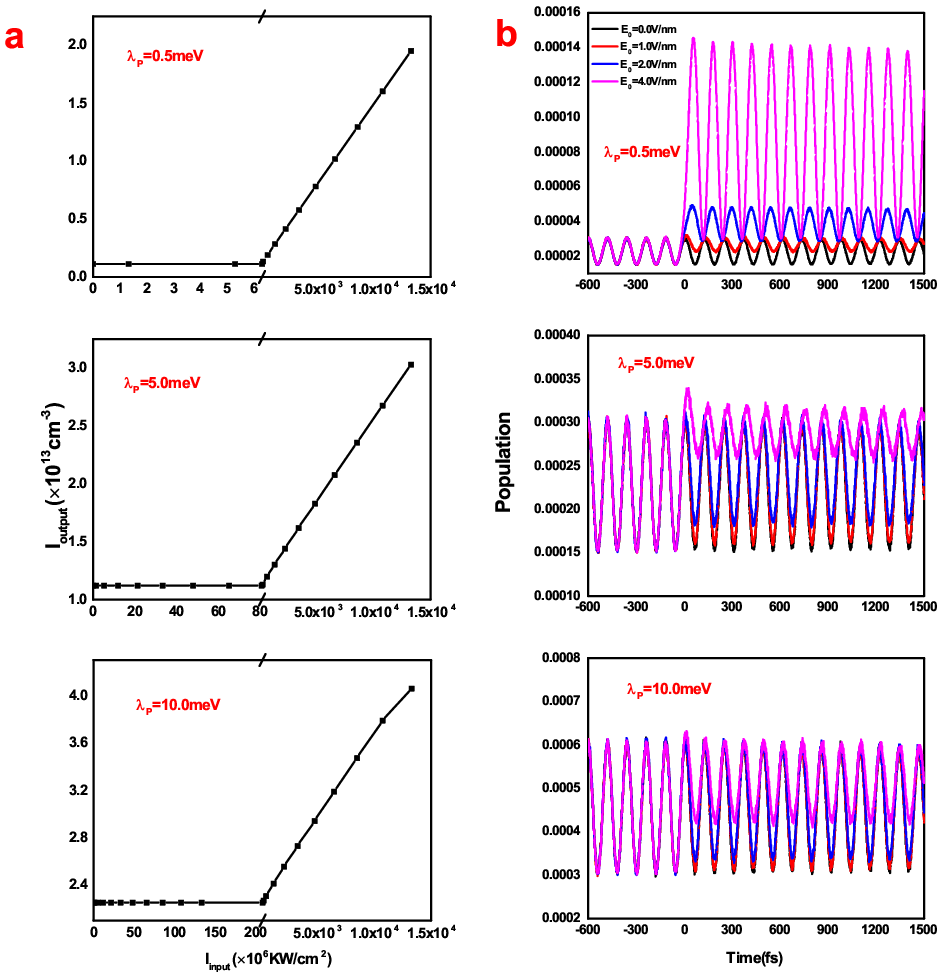}
  \caption{Panel a: The $I_{output}$-$I_{input}$ curve at different intracavity photon-outside
cavity bath coupling strength $\lambda_P$; Panel b: The population dynamics of photon state inside the subsystem including one-molecule and one-photon at different intracavity photon-outside
cavity bath coupling strength $\lambda_P$.}
  \label{fig5}
\end{figure}

Firstly, we calculated the structure-property relationship between the intracavity photon-outside
cavity bath coupling strength $\lambda_P$, cavity length $L_{cavity}$, and laser threshold $I_{thred}$. The
calculated results are shown in the Figure \ref{fig3}.
Figure \ref{fig3}a shows the structure-property relationship between the intracavity photon-outside
cavity bath coupling strength $\lambda_P$, cavity length $L_{cavity}$, and laser threshold $I_{thred}$, while
Figure \ref{fig3}b shows the population dynamics of photon state. Without loss of generality, we set that the quality of this order
of magnitude is often used for research in experiments, the $\lambda_P$ is $0.5~meV$. From the calculation results, it can be
seen that the laser threshold increases with $L_{cavity}$. The laser intensity threshold first
decreases and then increases,
which can be understood from physical picture. The external field will destroy
the equilibrium state of the system and bring it to a new stable state. Only when the intensity
of the external field $E_{pump}(t)$ is sufficient to break the equilibrium state of the system, lasing
phenomenon will happened, which manifest as the broken line of $I_{output}$-$I_{input}$ curve and
as a sudden change on the dynamics of intracavity photon state. When the coupling strength
between exciton and intracavity photon is large
enough, a stronger external field is required to disrupt the equilibrium state of the intracavity system;
And, when $L_{cavity}$ is very large, the population transfer between exciton and intracavity
photon is suppressed due to the small coupling strength between exciton and
the intracavity photon. Therefore,
a stronger external field is needed to produce a sudden change on the dynamics of intracavity photon state. From
this physical picture, it is not difficult to understand that the laser threshold varies with
$L_{cavity}$. It is worth noting that there is no direct correspondence between the laser threshold and the
steady-state population of the intracavity photon state. This is because the laser threshold is controlled by
the short-time excitation of the external field, and it belongs to the short-time dynamical property; the
steady-state population of the intracavity photon state is controlled by the interaction of different components
in total system, which belongs to the long-time dynamical property. So, the phenomenon of the low laser threshold
and the large steady-state population of the intracavity photon state may happen. This is indeed reflected
in Figure \ref{fig4}. The population dynamics
of ground state, local excited state and photon state at different
cavity length are shown in Figure S1, S2 and S3 (see the Supporting Information). Of course, when the
cavity length $L_{cavity}$ is very large (the exciton-photon coupling $V_{ep}$ is smaller
than the intracavity photon leakage rate $\Gamma_P$),
the steady-state population of the intracavity photon state will indeed show a monotonic decreasing
trend as $L_{cavity}$ increases, shown in Figure S4 (see the Supporting
Information). From Figure \ref{fig3}b, we can see that
the laser intensity threshold has a linear relationship as the
intracavity photon-bath coupling strength increasing. And, the laser intensity threshold is inversely proportional to the quality
of cavity. When the intracavity photon-bath coupling strength is set to $0.0$,
the laser intensity threshold is also equal to zero. The results are shown in Figure S7 (see the Supporting Information). This
indicates that when we change the the
intracavity photon-bath coupling strength values,
we can obtain laser thresholds of any size. Base on this linear relationship, we are not
limited to the specific values of the laser threshold, but focus on the structure-property
relationship. It is worth noting that as the intracavity photon leakage rate increases, it does
lead to an increase in the initial population of the intracavity photon state and a
decrease in the steady-state population of the intracavity photon state, shown in Figure \ref{fig5}.

\begin{figure}[ht]
 \centering
  \includegraphics [height=5cm,angle=0]{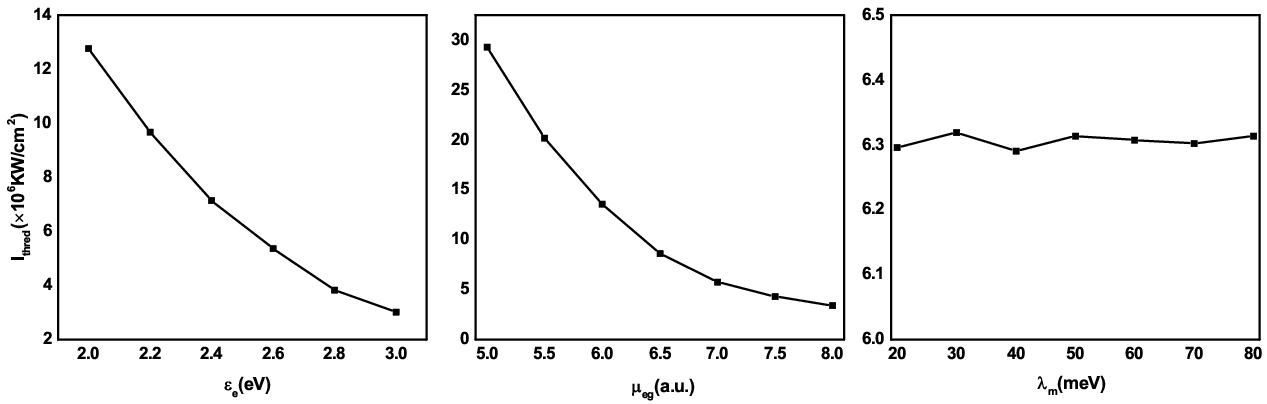}
  \caption{The structure-property relationship between the molecular $S_1$ state energy $\epsilon_e$, the molecular transition dipole $\mu_{eg}$ and the reorganization energy $\lambda_m$ and the laser threshold $I_{thred}$.}
  \label{fig6}
\end{figure}

\begin{figure}[ht]
 \centering
  \includegraphics [height=8cm,angle=0]{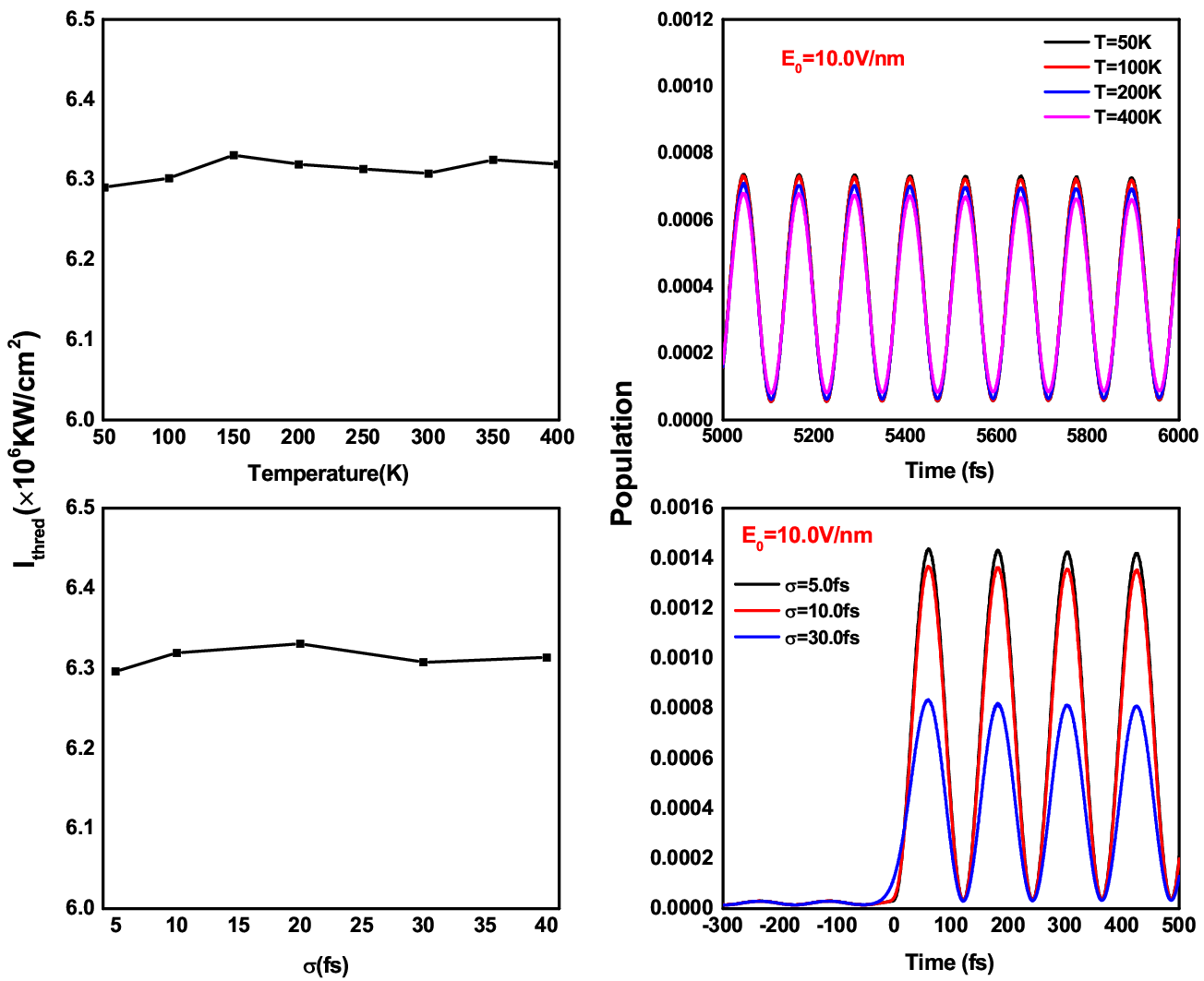}
  \caption{Panel left: The relationship between temperature $T$, duration $\sigma$ and the laser threshold $I_{thred}$; Panel right: The population dynamics of photon state inside the subsystem including one-molecule and one-photon at different temperature $T$ and duration $\sigma$.}
  \label{fig7}
\end{figure}

Next, we calculate the structure-property relationship between the electronic structure
properties of single molecule and the laser threshold. We investigate the influence of
the energy of $S_1$ state $\epsilon_e$, the molecular transition dipole $\mu_{eg}$ and the reorganization energy $\lambda_m$
on the laser threshold $I_{thred}$. Here, we set $L_{cavity}=15.0nm$, $\lambda_P=0.5~meV$. For the
calculation of reorganization energy, we use Debye spectral density
$J(\omega)=\frac{2\lambda_m\omega\omega_c}{\omega^2+\omega_c^2}$ to describe the exciton-phonon coupling. Here,
we change the reorganization energy $\lambda_m$. And, the characteristic frequency
$\omega_c$ is set to ~$1450.0$~cm$^{-1}$.\cite{berkelbach13114103,zang175105,zhang211654} The calculated results are shown in the Figure \ref{fig6}. From the
results, we can see that the laser threshold $I_{thred}$ are monotonically decreasing as the molecular
$S_1$ state energy $\epsilon_e$ and the molecular transition dipole $\mu_{eg}$ increasing. But, the
reorganization energy $\lambda_m$ cannot change the laser threshold $I_{thred}$.
The molecular $S_1$ state energy $\epsilon_e$ and the molecular transition dipole $\mu_{eg}$ are inversely proportional to the laser
threshold $I_{thred}$. This is consistent with the experimental results.\cite{oyama214112} When the molecular
$S_1$ state energy $\epsilon_e$ decreases, the intracavity photons will be quenched faster (shown in Table \ref{tab2}), which
leads to the increase of the laser threshold $I_{thred}$. The decreasing of the molecular transition dipole $\mu_{eg}$ will weaken the ability
of external fields to disrupt the initial steady-state of the intracavity system, leading to the increase of the laser threshold $I_{thred}$. Although
the reorganization energy $\lambda_m$ has almost no effect on the laser threshold $I_{thred}$. However, the increasing of the
reorganization energy $\lambda_m$ can reduce the steady-state population of intracavity photon state, which does
hinder light amplification in the cavity. The calculated population dynamics of intracavity photon state is shown in
Figure S8 (see the Supporting Information). As we know, Debye spectral density is very suitable for describing the low-frequency
vibrational modes of molecule. When we increase $\lambda_m$, the low frequency vibration modes are enhanced, and
the high-frequency modes are almost unaffected. Therefore, the molecular "four-level" energy system is ruined. This is consistent with the conclusion
proposed by Ref. \citenum{ma157004}. In this work, we cannot consider the influence of the molecular aggregation
effect on the laser threshold $I_{thred}$. And, we will extend the current work
to investigate the influence of the molecular aggregation
effect on the laser threshold $I_{thred}$ in our future work.

Finally, we
investigate the effects of temperature $T$ and external field duration $\sigma$ on the laser
threshold $I_{thred}$, and the calculated results are
shown in the figure \ref{fig7}. From the calculation results, we can see that temperature $T$ and
external field duration $\sigma$ cannot change the laser threshold $I_{thred}$. From the
population dynamics of intracavity photon state, it can be seen that
the increasing of temperature $T$ can accelerate the relaxation of the intracavity system. And,
the increasing of the external field duration $\sigma$ can decrease the external
field strength at $t=0.0$, thereby reducing the population of intracavity photon state. Although
the action time of the external field $E_{pump}(t)$ will increase as the duration $\sigma$ increasing, a
smaller duration $\sigma$ is indeed beneficial for increasing the steady-state population of intracavity photon
state. It is worth noting that increasing duration $\sigma$ does not mean that the external field
$E_{pump}(t)$ can exhibits the performance of continuous wave. when the duration $\sigma$ approaches
infinity, the external field $E_{pump}(t)$ has no effect on the intracavity system.

To conclude, based on the Time-dependent wavepacket diffusion (TDWPD) method in coupled with light-matter
interaction, we develop a microscopic quantum dynamic approach to  describe the organic lasing phenomena in dissipative cavity. The extended TDWPD is applied to investigate the structure-property relationships between the lasing threshold and the
intracavity photon-outside cavity bath coupling strength, cavity size, single molecular
electronic structure properties. The following conclusions
are drawn: (i) A microscopic model suitable for describing the lasing
dynamics of organic molecular system has been constructed, which can be used to describe
the structure-property relationships between laser
threshold and cavity parameters, molecular electronic structure parameters; (ii) The
microscopic physical picture of organic laser is proposed. The photons outside the
cavity can penetrate into the cavity, leading to the thermal equilibrium state in the cavity. Only
when the intensity of the external field is sufficient to break the
equilibrium state of the intracavity system, lasing
phenomenon will happened; (iii) The laser threshold decreases first and
then increases as the cavity size increases, and there is an optimal
value that can be understood from physical picture. The laser threshold
increases linearly with increasing the intracavity photon-outside
cavity bath coupling strength, exhibiting the monotone decreasing tendency
with the increase of the cavity quality value (Q value), which is consistent with
the experimental results; (iv) The reorganization energy cannot change the
laser threshold. The larger the reorganization energy leads to the reduction of the
steady-state population of intracavity photon state. The energy of $S_1$ state and
the transition dipole are inversely proportional to the laser threshold, which is
consistent with the experimental conclusion. The proposed formalism and
structure-property relationship is beneficial to understanding the mechanism of
organic laser and optimizing the design of organic laser materials.
\section*{ACKNOWLEDGEMENT}
Financial supports from Shenzhen Science and Technology Program and the National Natural Science
Foundation of China (Grant Nos. 21788102) as well as
the Ministry of Science and Technology of China
through the National Key R$\&$D Plan (Grant No.
2017YFA0204501) are gratefully acknowledged.

\begin{suppinfo}
The supplementary materials include the population dynamics
of ground state, local excited state and photon state inside
the subsystem including one-molecule and one-photon at different
cavity length, The $I_{output}$-$I_{input}$ curve at cavity length
$L_{cavity}=15.0$~nm and the intracavity photon-outside
cavity bath coupling strength $\lambda_P=0.0$~meV, the population
dynamics of photon state inside the subsystem including one-molecule
and one-photon at different reorganization energy $\lambda_m$.
\end{suppinfo}

\providecommand*\mcitethebibliography{\thebibliography}
\csname @ifundefined\endcsname{endmcitethebibliography}
  {\let\endmcitethebibliography\endthebibliography}{}

\end{document}